\documentclass[acmsmall,screen]{acmart}
\usepackage{amsfonts}
\usepackage{textcomp}
\usepackage{xcolor}
\usepackage{listings}
\usepackage{subfigure}
\usepackage{multirow}
\usepackage{colortbl}
\usepackage{xspace}
\usepackage{url}
\usepackage{enumitem}
\usepackage{titlesec}
\usepackage{booktabs}
\usepackage{algorithm}
\usepackage[noend]{algpseudocode}
\usepackage{rotating}
\usepackage[normalem]{ulem}
\usepackage{color}
\usepackage{pifont}
\usepackage{bbding}
\usepackage{svg}
\usepackage{threeparttable}
\usepackage{balance}
\usepackage{bm}
\usepackage{color}
\usepackage{transparent}
\usepackage{import}
\usepackage{hyperref}
\usepackage{cleveref}
\usepackage[many]{tcolorbox}
\usepackage{flushend}
\usepackage{subcaption}
\tcbuselibrary{listings,breakable}

\newcommand{\lida}[1]{\textcolor{black}{#1}}

\newcommand{\toolname}{JC-Finder\xspace}

\usepackage{tikz}

\usepackage{titlesec}
\usepackage{setspace}

\definecolor{codegreen}{rgb}{0,0.6,0}
\definecolor{codegray}{rgb}{0.5,0.5,0.5}
\definecolor{codepurple}{rgb}{0.58,0,0.82}
\definecolor{backcolour}{rgb}{0.95,0.95,0.92}

\lstdefinestyle{mystyle}{
    backgroundcolor=\color{backcolour},   
    commentstyle=\color{codegreen},
    keywordstyle=\color{magenta},
    numberstyle=\tiny\color{codegray},
    stringstyle=\color{codepurple},
    basicstyle=\ttfamily\footnotesize,
    breakatwhitespace=false,         
    breaklines=true,                 
    captionpos=b,                    
    keepspaces=true,                 
    numbers=left,                    
    numbersep=5pt,                  
    showspaces=false,                
    showstringspaces=false,
    showtabs=false,                  
    tabsize=2
}

\begin{document}
\title{\toolname: Detecting Java Clone-based Third-Party Library by Class-level Tree Analysis}


\author{Lida Zhao}
\authornote{Also with Singapore Management University.}
\affiliation{%
  \institution{Nanyang Technological University}
  \country{Singapore}
}
\email{LIDA001@e.ntu.edu.sg}

\author{Chaofan Li}
\affiliation{%
  \institution{Huazhong University of Science and Technology}
  \country{China}}
\email{chaofanli1@hust.edu.cn}

\author{Yueming Wu}
\authornote{Yueming Wu is the corresponding author.}
\affiliation{%
  \institution{Huazhong University of Science and Technology}
  \country{China}}
\email{yuemingwu@hust.edu.cn}

\author{Lyuye Zhang}
\affiliation{%
  \institution{Nanyang Technological University}
  \country{Singapore}
}
\email{zh0004ye@e.ntu.edu.sg}

\author{Jiahui Wu}
\affiliation{%
  \institution{Nanyang Technological University}
  \country{Singapore}
}
\email{JIAHUI004@e.ntu.edu.sg}

\author{Chengwei Liu}
\affiliation{%
  \institution{Nanyang Technological University}
  \country{Singapore}
}
\email{chengwei.liu@ntu.edu.sg}

\author{Sen Chen}
\affiliation{
  \institution{Nankai University}
  \country{China}
}
\email{senchen@nankai.edu.cn}

\author{Yutao Hu}
\affiliation{%
  \institution{Huazhong University of Science and Technology}
  \country{China}}
\email{yutaohu@hust.edu.cn}

\author{Zhengzi Xu}
\affiliation{%
  \institution{Nanyang Technological University}
  \country{Singapore}
}
\email{zhengzi.xu@ntu.edu.sg}

\author{Yi Liu}
\affiliation{%
  \institution{Nanyang Technological University}
  \country{Singapore}}
\email{yi009@e.ntu.edu.sg}

\author{Jingquan Ge}
\affiliation{%
  \institution{Nanyang Technological University}
  \country{Singapore}}
\email{jingquan.ge@ntu.edu.sg}

\author{Jun Sun}
\affiliation{%
  \institution{Singapore Management University}
  \country{Singapore}
}
\email{junsun@smu.edu.sg}

\author{Yang Liu}
\affiliation{%
  \institution{Nanyang Technological University}
  \country{Singapore}
}
\email{yangliu@ntu.edu.sg}

\thispagestyle{plain}
\pagestyle{plain}

\begin{abstract}
\lida{While reusing third-party libraries (TPL) facilitates software development, its chaotic management has brought great threats to software maintenance and the unauthorized use of source code also raises ethical problems such as misconduct on copyrighted code.} 
To identify TPL reuse in projects, Software Composition Analysis (SCA) is employed, and two categories of SCA techniques are used based on how TPLs are introduced: clone-based SCA and package-manager-based SCA (PM-based SCA). 
\lida{Although introducing TPLs by clones is prevalent in Java, no clone-based SCA tools are specially designed for Java. Also, directly applying clone-based SCA techniques from other tools is problematic.}
To fill this gap, we introduce \toolname, a novel clone-based SCA tool that aims to accurately and comprehensively identify instances of TPL reuse introduced by source code clones in Java projects. 
\toolname achieves both accuracy and efficiency in identifying TPL reuse from code cloning by capturing features at the class level, maintaining inter-function relationships, and excluding trivial or duplicated elements.
To evaluate the efficiency of \toolname, we applied it to 9,965 most popular Maven libraries as reference data and tested the TPL reuse of 1,000 GitHub projects. The result shows that \toolname achieved an F1-score of 0.818, outperforming the other function-level tool by 0.427. The average time taken for resolving TPL reuse is 14.2 seconds, which is approximately 9 times faster than the other tool. We further applied \toolname to 7,947 GitHub projects, revealing TPL reuse by code clones in 789 projects (about 9.89\% of all projects) and identifying a total of 2,142 TPLs. 
\toolname successfully detects 26.20\% more TPLs that are not explicitly declared in package managers.
\end{abstract}

\ccsdesc[500]{Software and its engineering~Reusability}
\keywords{Code Clone, Java, SCA}
\maketitle
\section{Introduction}
Third-party libraries (TPL) are widely adopted to facilitate software development, enhancing the efficiency of software development. To understand individual software components or modules that constitute a software system as completely as possible, Software Composition Analysis (SCA) has been widely adopted. However, the management of TPLs is often disorganized, with many TPLs not being introduced through package managers,  leading to poor version management~\cite{Kula2017}. This disorganization presents challenges to SCA tools and can result in significant ethical issues, including license violation and code plagiarism~\cite{osspolice2017}.

\lida{Specifically, software components are commonly imported explicitly through package managers~\cite{mavenintro,gradle}, a method widely supported by most SCA tools~\cite{owasp, steady, Snyk}. These tools are referred to as PM-based SCA tools.}
Alternatively, TPLs can also be introduced through the copy-and-paste of files, functions, or code snippets from other source code projects. According to Dejavu~\cite{Lopes2017}, code clones are common in Java source code projects where about 14\% projects have more than 50\% files overlapped in other Java projects.
\lida{Several clone-based SCA tools support detecting such TPLs. Snyk~\cite{Snyk} identifies TPLs by matching file paths or entire file hashes against known library files. CENTRIS~\cite{woo2021centris} detects TPLs based on matching function features of source code projects. 
However, none of these tools are specially designed for Java, neglecting essential Java features and leading to significant problems. For instance, Java is an Object-Oriented Programming (OOP) language where a class is the most fundamental unit~\cite{basicunit}. It embodies \textit{encapsulation}~\cite{encapsulation} by integrating data along with its corresponding functions. We found that some files contain multiple classes and only some of them are reused (Section \ref{sec:fileclone}), which makes Snyk's file-based detection method less effective~\cite{snykdocforc}. 
Furthermore, within OOP, the complexity of inter-function relationships significantly exceeds the simplicity of individual functions representing single functionalities. For example, \textit{polymorphism}~\cite{polymorphism} enriches the dynamic function binding, while \textit{inheritance}~\cite{inheritance} adds diversity to function implementation. These principles, along with the SOLID guidelines~\cite{martin2000design}—particularly the Interface Segregation Principles~\cite{ISP} and Dependency Inversion principles~\cite{DIP}—underscore the need to recognize the multiple connections among functions in order to truly capture comprehensive functionality. OOP design principles reveal that methods are not isolated; they exist in a network of collaboration and specialization. This complexity highlights that a mere focus on function similarity, without considering the higher-level functionality that these relationships facilitate, leads to incomplete inferences in TPL analysis.
Unfortunately, the function-based method of CENTRIS fails to maintain such relations, resulting in weak functionality representation. Furthermore, the overwhelming functions also causes Feature Redundancy (\Cref{sec:reprelimi}), where accounting for features of trivial or duplicated functions could introduce significant noise, leading to false positives and degrading the performance of the detector.} In total, we mainly address two challenges:
\lida{
\begin{itemize}[leftmargin=*]
\setlength\itemsep{0em}
    \item \emph{Challenge 1: No tool sufficiently addresses Java-specific features, such as treating the class as the fundamental unit of reuse or preserving inter-function relations, leading to suboptimal functionality representation.}
    \item \emph{Challenge 2: The considerable number of features from trivial functions introduces noise and slows down the detector.}
\end{itemize}
}

To fill this gap, we propose a novel tool named \textbf{J}ava \textbf{C}lone-based TPL \textbf{Finder} (\toolname) to accurately and comprehensively identify the TPLs introduced by code clones in Java source code projects.

To address the first challenges, we started by conducting experiments to study how the TPLs are being introduced and then determined the optimal granularity for Java clone-based SCA. The results indicate that functions within the same class tend to be imported together, showcasing significant inter-function relations. Class-level granularity strikes a balance by being less coarse than file-level while preserving more inter-function relations than function-level granularity. Consequently,  we selected class-level clone techniques for our SCA tool.
Regarding the second challenge, we employed a set of feature filtering rules, including the filtering of supporting classes (defined in \Cref{sec:supportclasses}), filtering classes based on centrality, and filtering duplicated classes through clone detection methods. These filtering techniques help to minimise noise, increase the accuracy of results and improve the efficiency of the tool.


To evaluate the effectiveness of \toolname, we collected a \textit{Reference Dataset} comprising 9,965 Maven libraries (in \Cref{sec:datacollection}). For comparison with function-level clone-based SCA tools, we chose CENTRIS~\cite{woo2021centris}. Although CENTRIS was originally tested in C/C++ projects, the authors asserted it is applicable to any programming language. Therefore, we created a Java version of CENTRIS, termed CENTRIS4J, which replicates the original tool's core logic but changes entrance from C/C++ to Java (in \Cref{sec:toolprepare}). 
Next, we compared the performance and efficiency of \toolname and CENTRIS4J on 1,000 GitHub projects. Furthermore, we compared the TPL reuse results obtained through \toolname with the list of dependencies declared through package managers. This allowed us to identify the number of TPL reuses that were missed by solely relying on package managers for SCA.
The experimental results revealed that \toolname achieves a recall of 0.986 and a precision of 0.698, outperforming CENTRIS4J by 0.462 and 0.353, respectively. Moreover, \toolname took 52 hours to generate features from the \textit{Reference Dataset}, which is about four times faster than that of CENTRIS4J, which took 220 hours. In terms of inferring TPL reuse, \toolname consumed an average of 14.2 seconds per project, which is about nine times faster than that of CENTRIS4J, which takes about 126.6s.
Furthermore, \toolname identified 26.20\% more TPLs introduced by code clones compared to the TPLs declared by package managers. This highlights the effectiveness of \toolname in complementing traditional Java PM-based SCA, which relies solely on package managers.

The main contributions are as follows:
\begin{itemize}[leftmargin=*]
    \item {We introduce \toolname, the first clone-based SCA tool specifically designed for Java and the first in applying class-level granularity among all clone-based SCA tools.} 
    \item We manually construct a ground truth of clone-based TPL reuse for 1,000 GitHub projects for public use. 
    \item \toolname achieves a recall of 0.986 and a precision of 0.698, outperforming the state-of-the-art tool by 0.280 and 0.428, respectively. Also, \toolname improves the PM-based SCA by locating 26.20\% more TPLs introduced by cloning.
\end{itemize}

\section{Related Work}

\subsection{Software Composition Analysis}
{SCA is a critical software auditing technique, primarily used for detecting TPLs to generate a Software Bill of Materials (SBOM). TPLs can be introduced in three ways~\cite{zhao2023}: by package managers, by external references, and by code cloning. }

{To detect TPLs introduced by package managers, tools such as Dependabot~\cite{dependabot}, OSSIndex~\cite{ossindex} parse the manifest files (e.g. pom.xml) of package managers for user-specified dependencies. Other tools like Eclipse Steady~\cite{eclipse}, and OWASP Dependency Check~\cite{owasp} can build projects and also resolve transitive dependencies, which are the dependencies automatically imported to support the direct dependencies. 
However, all of them rely on package managers and are unable to detect library reuse based on source code clone detection. }

{External references refer to the components relevant to the project but absent from the BOM, including individual components, services, or the BOM itself~\cite{externalref}. Some commercial tools, like snyk~\cite{Snyk}, support detecting these external references.}

TPLs can also be introduced by code clones through copying and pasting files or code snippets. Therefore, clone detection techniques are utilized to identify code similarity which further facilitates TPL recognition. Snyk CLI~\cite{Snyk} uses file-level clone detection. CENTRIS~\cite{woo2021centris} and OSSFP~\cite{ossfp} detect library reuse in C/C++ with function-level clone detection. 

\subsection{Code Clone Detection}
{Code clone detection is a technique to identify code similarity. When detecting TPLs introduced by copy-and-paste, the clone technique serves as the upper stream procedure for SCA, followed by more procedures such as feature extraction, feature filtering, etc. 
Previous studies~\cite{Rattan2013,Ain2019,Bellon2007} have defined 4 types of clones. Type-1 Clones are code snippets that are almost identical in text, differing only in spaces, blank lines, and comments. Type-2 Clones include code snippets that are the same except for changes in identifiers like function names, class names, and variables. Type-3 Clones involve snippets that vary at the statement level, with possible additions, modifications, or deletions of statements. Type-4 Clones are code snippets that, although syntactically different, perform identical functions and thus show functional similarity. Our technique handles Type 1/2/3 clones, allowing for modified variable names and reordered code lines by applying AST and order-insensitive hash. In addition, TPL reuse always indicate code clones but the presence of clones does not necessarily indicate determined TPL reuse, as the classes might be support classes that do not clearly represent TPL reuse.

To better recognize code clones, different granularity is utilized. The applicable granularity of clone detection includes file-level, class-level, and function-level.}

For file-level code clone detection techniques \cite{johnson1994substring, gode2009incremental, jadon2016code, kamiya2002ccfinder, patenaude1999extending, ragkh2018apicture, haque2016generic}, they compare entire source code files to determine if there is any similarity between them. 
For example, Johnson et al. \cite{johnson1994substring} detect clones using the fingerprint-matching method. 
CCFinder \cite{kamiya2002ccfinder} extracts token sequences from program code and utilizes transformation rules to convert the token sequence, enabling the detection of Type-1 and Type-2 clones.

For class-level code clone detection techniques \cite{tessem1998retrieval, tamada2004design, basit2005detecting, sager2006detecting, sager2006code, chilowicz2009syntax},
Tessem et al. \cite{tessem1998retrieval} describe the methods used in the retrieval phase of a case-based component in a prototyping tool for the Java programming language. 
Sager et al. \cite{sager2006detecting} present a novel approach to detect similarities between different Java classes based on abstract syntax trees. 
Similarity is calculated by means of three tree similarity algorithms: bottom-up maximum common subtree isomorphism, top-down maximum common subtree isomorphism, and the tree edit distance.

For function-level clone detection techniques \cite{ragkh2017using, kim2018software, li2017cclearner, wang2018ccaligner, golubev2021multi, hung2020cppcd, jiang2007deckard, wei2017cdlh, zhang2019astnn, liang2021astpath, jo2021twopass, hu2022treecen, wu2022amain, wang2017ccsharp, zhao2018deepsim, wu2020scdetector, zou2020ccgraph, saini2018oreo, hua2021fcca}, Deckard \cite{jiang2007deckard} utilizes Locality Sensitive Hashing (LSH) to detect clones by clustering similar vectors computed from Abstract Syntax Trees (ASTs) of any language with grammatical regulations.
CCGraph \cite{zou2020ccgraph} reduces the Program Dependency Graph (PDG) size to improve efficiency. 
It then filters the PDG using numerical features for coarse filtering and string similarity for a second filtering step within each category. 
Oreo \cite{saini2018oreo} utilizes a combination of machine learning, information retrieval, and software metrics to detect clones with high precision and recall, showing the potential of integrating multiple techniques to improve performance.


\section{Preliminary Study}\label{sec:prestudy}
\lida{In this section, we aimed to find the optimal granularity for clone-based SCA. Typically, existing clone detection works are performed at three main levels: file-level~\cite{johnson1994substring, gode2009incremental, jadon2016code}, class-level~\cite{sager2006detecting, sager2006code, chilowicz2009syntax} and function-level~\cite{ragkh2017using, kim2018software, li2017cclearner}. Initially, we evaluated the efficacy of file-level and class-level clones in recognizing TPLs, revealing the inherent limitations in file-level analysis with examples. Subsequently, we investigated the code clone conditions at the function level, as employed by leading SCA tools~\cite{woo2021centris, ossfp}. By introducing the concepts of ``associate clone'' and ``conjugate clone'', we highlighted the inadequacies of relying solely on function-level granularity for TPL identification. Ultimately, we chose class-level granularity for subsequent experiments.}

\subsection{Data Collection}
For this preliminary study, {we randomly collected 1,000 Java GitHub projects with more than 20 stars, excluding forks, examples, and educational projects.} 

\subsection{File-level Clone and Class-level Clone}\label{sec:fileclone}
{In our experiment, we first investigated the general clone condition with hash-matching for both file-level and class-level clones. Then, we highlighted the limitations of file-level cloning with an example of class-level code reuse.}

For file-level clones, we first discarded empty or configuration files, leaving 179,182 files. We then normalized the content of these files by eliminating all comments and quoted strings. Additionally, we removed the package information at the beginning of each file. Afterward, all spaces and new lines were stripped out before calculating the hash value for each file. Our analysis revealed that 7,457 files, approximately 4.16\% of the total, had identical hash values when compared to other files in the dataset. 
For class-level clones, we utilized JavaParser~\cite{javaparser} for class partitioning. Note that JavaParser (version 3.25.5) has limitations with some newer Java features, resulting in unsuccessful parsing of certain Java files. Overall, 164,129 classes were identified from 134,710 files. Then, we applied the same normalization process used for file-level cloning to these classes. We found 9,048 classes (around 5.51\%) are cloned with other classes. Notably, 13,977 out of 134,710 files (10.38\%) were found to contain more than one class. {In such cases, if only some of the classes are copied, file-level cloning may fail to detect the code clone. 
Therefore, class-level hash clones result in a higher clone percentage than file-level hash clones.} To make it more intuitive, \textit{net.sourceforge .plantuml:plantuml} is an open-source UML (Unified Modeling Language) diagramming tool. It includes a file named \textit{AnimatedGifEncoder.java}, which consists of three classes: \textit{AnimatedGifEncoder}, \textit{NeuQuant}, and \textit{LZWEncoder}. While a GitHub project namely \textit{Reading-eScience-Centre/edal-java} \cite{edaljava} clones only one of the classes \textit{LZWEncoder}, file-level cloning is unable to detect this clone. As a result, class-level proves to be a more appropriate granularity for detecting TPL reuse in Java source code compared to file-level clones.

\subsection{Function-level Clone}\label{sec:funclevelclone}
Using TACC~\cite{wang2023comparison}, a recent scalable code clone detector, we identified a total of 350,636 functions from the 1,000 projects. Out of these functions, 197,769 were discovered to have clone relations with other functions, exhibiting a similarity of at least 0.7 (by default). These clone-related functions represent approximately 56.40\% of all the functions that were analyzed.
\lida{Although function-level detection uncovers more clones compared to file or class levels, it proves to be inadequate for recognizing TPL reuse in Java due to two main reasons. Firstly, function-level techniques overlook the inter-function relations, which is crucial in Java where \textit{polymorphism} and \textit{inheritance} are prevalent. \textit{Polymorphism} enables Java methods within the same class to perform varied functions based on the input parameters, while \textit{inheritance} allows a subclass to adapt and extend a method inherited from its superclass, enriching the runtime behavior. Preserving these intricate relationships is essential to recognizing the unique functionality offered by different implementations of similar functional signatures. Secondly, the inclusion of trivial functions lead to an overload of redundant features, increasing the risk of false positives.}

\subsubsection{Inter-function Relations}\label{sec:classclone}
\lida{To reuse a comprehensive functionality, a group of related functions is usually cloned together. However, the absence of inter-function relations can compromise the ability to infer the complete functionality, increasing the risk of inaccurate TPL recognition. To gain a deeper understanding of clone behavior,} we investigated two inter-function relations, including caller-callee relations (associated clone) and the co-existing relations (conjugate clone). \lida{The inter-function relations are studied within classes, because classes are fundamental units of encapsulating functionalities in OOP~\cite{basicunit}, providing insights into the design principles and modularity of the system~\cite{ooprograming}.}

\begin{figure}[ht]
    \centering
    \begin{minipage}{0.4\linewidth}
        \centering
        \includegraphics[width=\linewidth]{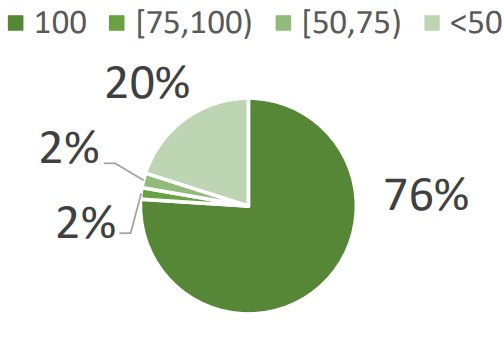}
        \captionof{figure}{Associated Clone \\Percentage Distribution}
        \label{fig:associateclonepercentdist}
    \end{minipage}%
    \begin{minipage}{0.4\linewidth}
        \centering
        \includegraphics[width=\linewidth]{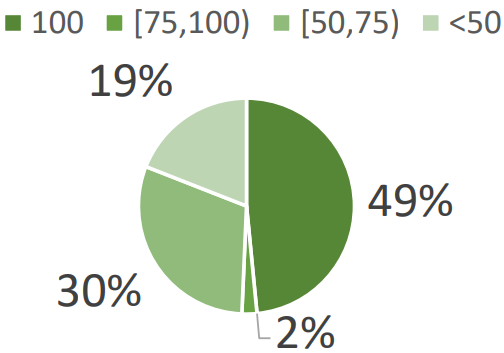}
        \captionof{figure}{Conjugate Clone \\Percentage Distribution}
        \label{fig:conjclonepercentdist}
    \end{minipage}
\end{figure}

{We first examined whether caller functions are cloned together with their callees. Such caller-callee function pairs are called \textit{associated clone pairs}. We started by mapping the caller-callee relations of all functions within each class. Then, we applied TACC~\cite{wang2023comparison} to resolve the function clone relations. We filtered the trivial functions (quantified definition in \Cref{sec:supportclasses}) and the clone relations within the same project. Next, we identified callee functions within the same class for each caller function and examined whether these callees were cloned alongside their respective callers. The proportion of cloned callees to total callees is named \textit{associated clone percentage}. \Cref{fig:associateclonepercentdist} shows that most of the caller (80\%) has a associated clone percentage higher than 50\%. Impressively, 76\% of the caller exhibit a 100\% associated clone percentage, meaning all their callees are associated cloned. }

\begin{figure}[tb]
    \centering
    \includegraphics[width=0.6\textwidth]{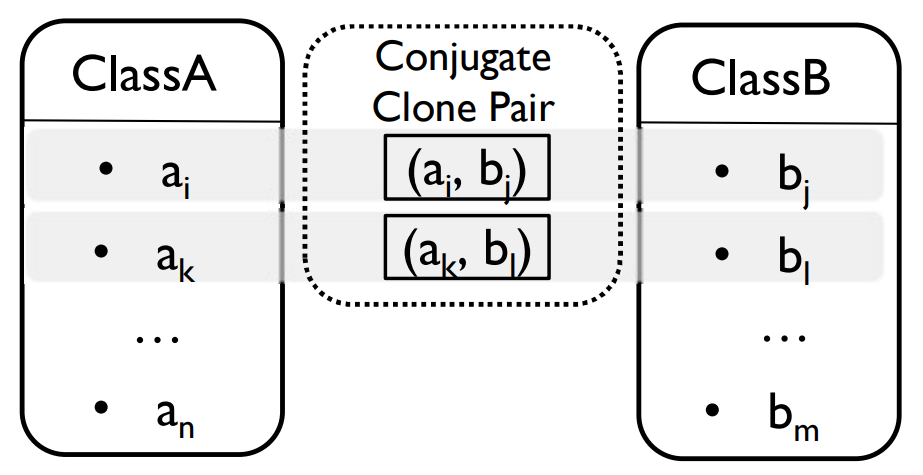}
    \caption{Conjugate Clone Pairs}
    \label{fig:conjugate-demo}
\end{figure}

{The analysis of associated clones confirmed that most callers are cloned alongside their callees within a class. Still, some functions can be cloned together even without a direct caller-callee relation. We introduced \textit{conjugate clone pairs} and conducted more experiments to dive deeper.}
\textit{Conjugate clone pairs} are clone pairs that occur between two classes, where multiple pairs of identical or similar functions are present. More formally, given two classes $A$ and $B$ with $n$ and $m$ functions respectively, let $C_{A,B}$ be the set of clone pairs between them. For any pair of clone functions $(a_i,b_j), (a_k,b_l) \in C_{A,B}$, if $i \neq k$ and $j \neq l$, then $(a_i,b_j)$ and $(a_k,b_l)$ are considered conjugate clone pairs. This deliberate design ensures one-to-one clone relationships between functions. It mimics the behavior of developers who, when deciding to copy a function, focus solely on the function they are copying without knowledge of how many other functions are similar. Conjugate clone pairs encompass associated clone pairs and extend to capture function interaction relations beyond just the caller-callee relationship.
\textit{The conjugate clone percentage} for $A$ and $B$ is then calculated as the ratio of the number of functions in conjugate clone pairs to the total number of functions in both classes, i.e.,

$$
\text{P} = \frac{\left\vert {(a_i,b_j), (a_k,b_l) \in C_{A,B} \mid i \neq k \text{ and } j \neq l } \right\vert * 2}{n + m}
$$
where $P$ is conjugate clone percentage and $\left\vert C_{A,B} \right\vert$ denotes the cardinality of set $C_{A,B}$.
An example of calculating the conjugate clone percentage is shown in \Cref{fig:conjugate-demo}. Suppose there are two classes, \textit{ClassA} and \textit{ClassB}, and both have 10 functions. There are only two pairs of clone functions ($a_i$, $b_j$) and ($a_k$, $b_l$) between them. Those two pairs are considered conjugate clones, and the conjugate clone percentage for \textit{ClassA} and \textit{ClassB} is calculated as $4 / (10 + 10)=0.2$.
Intuitively, more conjugate clone pairs indicate that the two classes may have been developed in a similar manner, possibly leading to library reuse.

Consequently, we determined the conjugate clone percentage with the following steps. Initially, we employed  TACC~\cite{wang2023comparison} to detect pairs of clone functions. {We excluded trivial functions, such as getter and setter, to minimize the influence of noise.}
For every pair, we traced back to the two classes related to the pair of functions to obtain two lists of functions. Subsequently, for every function in both lists, we checked if they had clone relations with any function in the other list. We recorded the clone pairs between the two classes and then used them to calculate the conjugate clone percentages. 


{The results are displayed in Figure \ref{fig:conjclonepercentdist}. A greater proportion of conjugate clones suggests a stronger inclination for functions to be duplicated alongside other related functions. Approximately 81\% of classes exhibit a conjugate clone percentage exceeding 50\%, and 48\% of these are entirely cloned, possessing a conjugate clone percentage of 100\%. The findings indicate that functions tend to be cloned alongside other functions in the same class rather than independently. This widespread phenomenon should not be overlooked during feature generation. To preserve this relation, class-level clone techniques are required for Java clone-based SCA.}

\subsubsection{Feature Redundancy}\label{sec:reprelimi}
A considerable amount of trivial functions, such as supporting functions like getters, setters, as well as \lida{utility functions implemented from superclasses through \textit{interitance}} such as \textit{toString()}, \textit{equals()}, and \textit{hashCode()}, could bring much noise and produce false positives for TPL identification. More detailed definitions are in \ref{sec:supportclasses}. To understand this intuitively, let us take a GitHub project \textit{Reading-eScience-Centre/edal-java}~\cite{edaljava} (\textit{P1}) as an example. The project deals with the manipulation and visualization of environmental data. {Suggested by the state-of-the-art tool, CENTRIS~\cite{woo2021centris}, we first detected clones between functions and then calculated the percentage of identical functions. The higher the percentage is, the more likely a library is cloned.} Using TACC, we examined \textit{P1}'s function clone relation with a Maven library called \textit{org.infinispan:infinispan-embedded} (\textit{L1}). Despite \textit{L1} being a Java distributed key-value storage system without any dependency relations with \textit{P1}, we found that out of 1,066 functions in \textit{P1}, 417 (about 39.1\%) had clone relations with the functions in \textit{L1}. However, most of these functions were found to be trivial functions. These functions introduced a considerable amount of noise, and none of them were representative enough to indicate TPL reuse.
{In contrast, we observed that \textit{P1} reused another Maven library, \textit{edu.ucar:netcdf}~\cite{netcdf} (\textit{L2}), across multiple classes, including \textit{DayOfMonthOfFixedYearDateTimeField} and \textit{MonthOfFixedYearDateTimeField}. Yet, the percentage of reused functions compared to those in \textit{L2} was only 20.09\% (223 out of 1067 functions). In this scenario, when a tool employs a threshold for TPL recognition (e.g., CENTRIS~\cite{woo2021centris}), regardless of threshold adjustments, there will inevitably be false positives or false negatives. To reduce these inaccuracies, it is necessary to exclude trivial functions and avoid using an ad-hoc threshold.} 

\begin{tcolorbox}[boxrule=0.5pt,arc=1pt,boxsep=-1mm,breakable]
    \textbf{Summary:} 
    {In detecting TPL via code cloning, class-level granularity offers a balanced approach, avoiding the under-representation and noise of function-level granularity and the excessive coarseness of file-level granularity.}
\end{tcolorbox}

\section{Methodology}\label{sec:methodology}

\begin{figure}[t]
    \centering
    \includegraphics[width=0.6\textwidth]{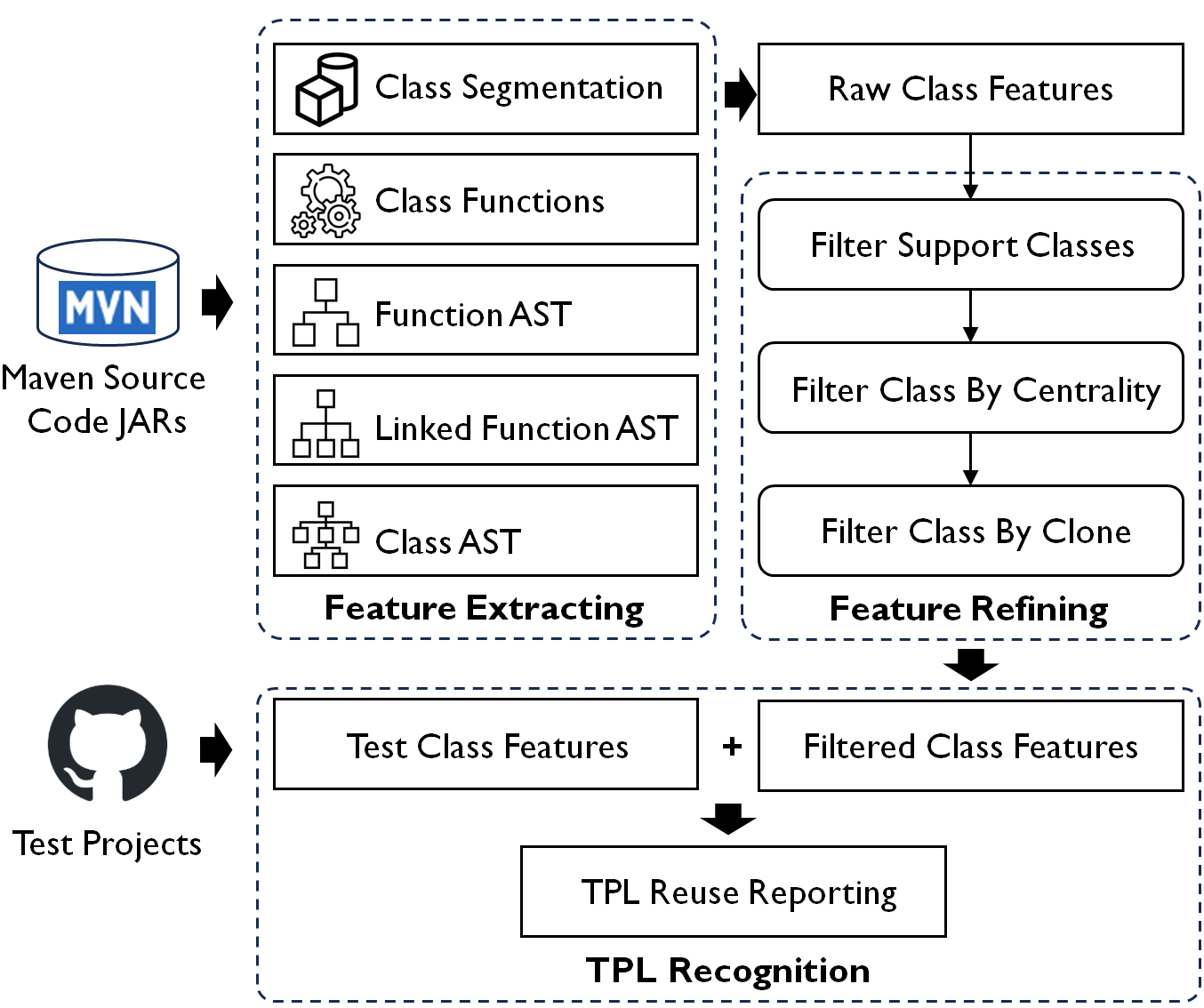}
    \caption{Overview of \toolname}
    \label{fig:toolOverview}
\end{figure}

\toolname comprises three main parts: Feature Extracting, Feature Refining, and TPL Recognition. Feature Extracting generates features for classes, Feature Refining refines the class features, and TPL Recognition is responsible for identifying potential source code clones and infers TPL reuse. An overview is presented in \Cref{fig:toolOverview}.

\subsection{Feature Extracting}
According to \Cref{fig:toolOverview}, extracting class features involves several steps. Firstly, we group the functions declared under the same class and normalize the code by removing comments and formatting. 
Next, We generate the Abstract Syntax Tree (AST) for each function using JavaParser~\cite{javaparser}, which is a syntax analyzer for Java. {In this process, inspired by Sager et al.~\cite{sager2006code, sager2006detecting}, we only keep the skeleton structure of the function and replace the concrete names or values with their corresponding types.} For example, all variable names are replaced as the generic string ``Simple Name'', and their values are replaced with basic types such as ``Primitive Type'' (for Java primitive types including int, bool, etc.), ``Literal'' (for all string values) and some customized class names. Such a technique makes the AST robust even if the variable name or values are changed. Subsequently, \lida{to maintain the inter-function relations}, we establish links between the function ASTs for functions within the same class based on the function invocations. Note that only internal function (functions declared under the same classes) invocations are linked and external function (functions declared in other classes) invocations are not linked. To achieve this, we locate and replace the function invocations with its function body recursively until all internal function invocations have been replaced. For external function invocations, we replace them with a placeholder node called ``Dummy External Node'' to indicate that we do not need to consider the content of these functions. The function invocation chains may form loops and we break them by prioritizing functions that invoke more other functions and have a greater number of lines of code (LOC). For example, if two functions, A and B, invoke each other, and A has more LOC than B, we only keep the invocations from A to B and break the invocation from B to A. Also, for self-recursive functions, we replace their function body with a ``Dummy Recursive Node'' to avoid infinite recursive loops.
After resolving all function invocations, we obtain the ``Linked Function AST''. A class can have multiple Linked Function ASTs, each starting with a specific function node. By gathering the starting function nodes and creating a class root node, we construct the ``Class AST''.

To facilitate similarity comparison, we need to derive the hash value of ``Class AST''. Although all ``Linked Function ASTs'' are attached under the ``Class AST'', their order should not affect the hash value. Therefore, similar to the previous work~\cite{lazar2014clone}, we adopted the following hash calculation method: (1) We start by calculating the hash value of all leaf nodes in the ``Class AST''. (2) Then, for each parent node of these leaf nodes, we calculate the hash value by adding its own hash value to the hashes of all its child nodes. (3) We repeat step (2) recursively until all nodes are visited. The hash value of the root node represents the hash value of the entire ``Class AST''. So, even if some code lines are rearranged or code blocks are re-ordered, the ``Class AST'' hash will remain unchanged.


\subsection{Feature Refining}\label{sec:featurerefining}

We perform Feature Extracting from 543,286 versions of the 9,965 Maven libraries (collected in \Cref{sec:datacollection}). Some of the features are duplicated and trivial which not only hampers efficiency but also leads to inaccurate recognition. For example, for the class reused in multiple TPLs, distinguishing the original one from the cloned ones is of vital importance. To make the features concise and meaningful, we implemented three filtering steps, including filter supporting class, filter classes by centrality, and filter classes by clone. 

\subsubsection{Filter Supporting Classes}\label{sec:supportclasses}
The first step is to remove supporting classes within a library. In a Java project, supporting classes refer to the classes that offer utility functions or data records, supporting the main algorithms and business logic. To effectively identify and filter out these supporting classes, we apply four specific criteria: C1: Interface and empty classes; C2: {Trivial functions within classes}; C3: Structural classes in design patterns; and C4: Test classes.

C1 removes interfaces without any concrete function implementations. Interfaces contain no function or a collection of function signatures, which compel their subclasses to implement all the declared functions. These classes define application functionalities as a contract, while the specific behaviors are to be implemented in other concrete classes. Therefore, they can be removed.

{C2 removes trivial functions within classes.} We noticed that some classes contain a group of extremely simple functions, such as getters, setters, and initializers, which provide basic functionalities in just a few lines of code. {They are considered noise in class feature extraction and should be removed.} We apply Maintainability Index (MI)~\cite{mi} to quantitatively identify the complexity of the functions and help distinguish the trivial functions. The MI is typically used to evaluate the maintainability of a program, based on three metrics: lines of code (LOC), cyclomatic complexity (CC), and Halstead volume (HV). It is calculated using
$$
MI = 171-5.2 \times HV - 0.23 \times CC - 16.2 \times LOC
$$
where MI is negatively correlated with program complexity. Thus, we evaluate the complexity of a function by adding the absolute values of the last three terms in the aforementioned equation. A higher resultant value indicates a more intricate function. For our analysis, we randomly picked 400 functions and identify their functionalities. We observed that all functions with a complexity lower than 60 provide simple functionalities with a few lines, such as assigning values, getters or setters. These examples of code complexity are illustrated in \Cref{fig:codecom}. Consequently, we consider functions with a complexity of less than 60 as trivial. {After removal, most classes retain their primary logical functions. However, some classes, primarily serving as data structure models and consisting only of trivial functions, are consequently removed.}


\begin{lstlisting}[language=Java, caption=Examples of Code Complexity, label=fig:codecom]
//code complexity: 63
private static void copyAllJars(String outputdir) {
  File out = new File(outputDir);
  string targetPath = out.getAbsolutePath()+"lib";
  File current = new File(".");
  string path = current.getcanonicalpath();
  string libPath = path + File.separatorChar+"lib";
  utils.copyFolder(libPath, targetpath, "jar");
}
// code complexity: 52
public MParam newParam(string name, string type) {
  MParam p = new MParam();
  p.setName(name);
  p.setType(type);
  return p;
}
\end{lstlisting}

C3 removes structural classes used in design patterns. Software design patterns are general, reusable solutions to commonly occurring problems within specific contexts in software design~\cite{designpatternwiki}. Some of the structural classes help organize the design patterns but do not contain concrete algorithms. For example, Abstract Factory~\cite{abstract-factory} requires a factory class to distribute registered objects, while Convertor and Adaptor~\cite{adapter} use convertor or adaptor classes to facilitate object transfer, etc. To achieve this, we reviewed all the design patterns from \cite{designpatternbook} and summarized a list of special naming patterns associated with structural classes in design patterns. We then filtered out these classes by matching their names accordingly. 

C4 removes test classes. If the paths or names of the classes matched with the terms ``test'' as suffix or preffix, we exclude them from the feature list.

\subsubsection{Filter Classes By Centrality}
In this stage, we focus on the interactions between classes and use \textit{PageRank}~\cite{pagerank} to measure the centrality of the classes. \textit{PageRank} measures the importance of a graph node based on the links pointing to it. Initially, we construct a class dependency diagram for all classes within a library based on the dependence relations between classes. For example, if \textit{ClassA} is used in \textit{ClassB}, we create a dependency edge from \textit{ClassA} to \textit{ClassB}. Generally, the \textit{PageRank} of a class is positively related to two factors: the number of classes it relies on and the importance of those classes. Note that \textit{PageRank} is a relative score within the class dependency diagram. To enhance interpretability, we convert it into percentiles, ranging from 0\% to 100\%, where 0\% is assigned to the top-scoring class, signifying its utmost importance in the library.

\begin{figure}
    \centering
    \includegraphics[width=0.8\textwidth]{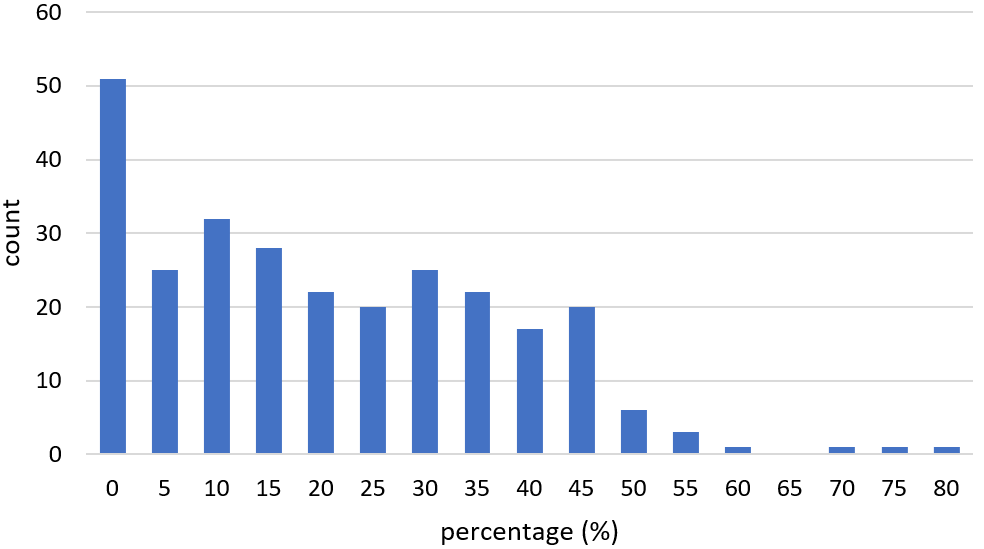}
    \caption{Percentile Distribution of Reused Classes}
    \label{fig:pagerankdist}
\end{figure}

During the collection of ground truth, we scanned 1,000 GitHub projects against all class features of the 9,965 libraries from the \textit{Reference Dataset} and manually examined whether the cloned classes indicate the TPL reuse (details in \Cref{sec:gtcollection}). Among all the classes that are confirmed to indicate TPL reuse, we picked 385 pairs of cloned classes to observe the distribution of their percentiles. The choice of 385 pairs was based on statistical considerations. It represents the minimum number of clone pairs required to achieve a confidence level of 95\% that the real value is within ±5\% of the results~\cite{statcalcu}. From the results shown in \Cref{fig:pagerankdist}, 95.63\% of classes are within the top 50\% percentiles. Therefore, we remove the classes within the bottom 50\% percentiles, as they are statistically considered to be supporting classes that are not representative enough to indicate TPL reuse effectively.

\subsubsection{Filter Classes By Clone}\label{sec:filterbyclone}
In the remaining classes, there are still numerous duplications, where different TPLs may contain the same class. For example, both \textit{LibA} and \textit{LibB} contain a \textit{ClassC}, making it difficult to tell whether \textit{LibA} copies \textit{ClassC} from \textit{LibB}, \textit{LibB} copies from \textit{LibA}, or both TPLs copy \textit{ClassC} from other libraries. Identifying the origin of \textit{ClassC} is crucial, otherwise, it could indicate multiple TPL reuse and increase the false positive TPL reuse results. To determine the originality of a class, we utilize the timestamps from the Maven libraries collected in \Cref{sec:datacollection}. The class with the earliest timestamp is considered the original class.
The duplication removal comprises two steps. The first step is to merge class features with the group names. The reason for operating within groups instead of artifacts is that the artifacts with the same group name are usually developed within a single source code project. It is common for them to share some code, and they are usually released together having the same timestamp. In this step, we merge the features with the same hash value and attach their ``artifact-version'' pairs to the feature. The timestamp of the feature is set as the timestamp of the earliest ``artifact-version'' pair. 
The second step is to remove feature duplication between groups. If two features are found duplicated, we only keep the earliest feature and discard the feature ``artifact-version'' list of the later feature. 

Finally, all features are unique and mapped with their original group names and list of artifacts-versions (GAV).

\subsection{TPL Recognition}
The TPL recognition process starts with extracting class features from the test projects. Next, we compare each of these features with the refined features obtained from the \textit{Reference Dataset}. Any matched features are then recorded along with their related GAV list. These matched features serve as evidence of TPL reuse. If there are one or more pieces of clone evidence associated with a particular TPL, we conclude that the TPL is reused by the project. As the final outcome, we generate a list of TPLs as the result.

\section{Evaluation}\label{sec:evaluation}
This work aims to provide a comprehensive understanding of TPL reuse through code clones in source code projects by introducing a new Java SCA tool based on class-level clone detection.
We focus on the following research questions:

\noindent \textbf{RQ1: Effectiveness Evaluation}: 
To answer the question, we first collected the ground truth of 1,000 source projects. We then conducted scans and evaluated the effectiveness of the tools using precision, recall, and F1-score as metrics.

\noindent \textbf{RQ2: Efficiency Evaluation}: 
This section compares the efficiency of \toolname with other clone-based SCA tools in Java.

\noindent \textbf{RQ3: TPL Detection Enhancement Analysis}: This research question studies the enhancement brought by \toolname compared to traditional PM-based SCA tools. How many projects reuse TPLs through code clone? How many of these TPLs are not declared through package managers? How many of these TPLs are already declared in the package managers? 


\subsection{Data Preparation}
This part details the steps of data collection and ground truth labeling.
\subsubsection{Data Collection}\label{sec:datacollection}

\textbf{Reference Dataset:} To obtain the library features for TPL reuse identification, we collected a reference dataset from the Maven Central Repository (MCR) \cite{mavenrepo}. We first used Libraries.io \cite{librariesio} to identify the most popular libraries in MCR based on their star numbers, and Libraries.io has a query limit of returning only 10,000 libraries per query. To avoid duplication of the source code projects, we then filtered the libraries built for different platforms (e.g., \textit{arm64}, \textit{x86}) or operating systems (e.g. \textit{darwin}, \textit{win}, \textit{linux}), we narrowed down our dataset to 9,965 libraries. For each selected library, we collected all their history versions and the corresponding releasing timestamps. In total, 543,286 versions were collected. Since we exclusively analyzed the original source code, we only analyzed the ``source-jar'' files, which did not include third-party libraries imported through the package manager.

\noindent \textbf{Test Dataset:} We collected this dataset to examine whether there are any unclaimed TPL reuses in real Java source projects. We searched for Java projects and sorted them based on their star numbers. However, due to GitHub's limitation, we could only retrieve the first 1,000 projects for each star number. We removed fork projects and tutorial projects. Finally, we collected 13,000 projects with star numbers ranging from around 70,000 to 20. Note that some of the \textit{Reference Dataset} originated from the GitHub projects, and we did not remove them. Then, we tried to build them by command \textit{mvn compile} to see if the project is intact. At last, 7,974 projects remained. 

\noindent \textbf{Test Data with Ground Truth:} This dataset is extracted from the \textit{Test Dataset} with elaborated ground truth for performance evaluation of tools. \lida{We randomly selected 1,000 projects from the \textit{Test Dataset} and verified the ground truth.} The subsequent section will elaborate on the steps of collecting ground truth. 

\subsubsection{Ground Truth Collection}\label{sec:gtcollection}
Identifying library reuse based solely on similarities is inadequate. To validate library reuse, we examined additional meta information such as class paths, change logs, copyright, license, and author list. \lida{It took 2 open-source software experts 2 months to label the ground truth.} The steps to establish the ground truth are as follows. 

\noindent \textbf{Paths \& Names:} First, we examined the paths and names of the classes. If the vendor name of the test project matched that of a library, we classified it as library reuse. For example, the original code repository of the library \textit{io.vertx:vertx-service-discovery} is \textit{vert-x3/vertx-service-discovery}. In addition, if the path to the test project contains library information, it indicates that the TPL is reused. For example, the class \textit{YieldingWaitStrategy} in the test project \textit{xiaoguichao/mr} had a path of \textit{/src/main/java/com/lmax/disruptor/}, indicating code reuse from the library \textit{com.lmax:disruptor}. Some library names are placed in a particular folder, such as \textit{/shared}, \textit{/ext}, etc.

\noindent \textbf{Change Logs:} Then, we checked the change logs in the class file. Some of the test classes were explicitly labeled with their origins using expressions like ``forked/copied from ...'', ``This class is a modified version of / based on the work of...'', etc. 

\noindent \textbf{Copyright \& License:} Next, we assessed the copyright and license information for both classes. We identified library reuse only when the test class lacked its own copyright/license, and the library class had its own copyright/license. In essence, we excluded cases where the test class had its own copyright/license, the library class lacked copyright/license information, or the library class had no copyright/license. 

\noindent \textbf{Author List:} Even if the classes possess copyright/licenses, they may not explicitly state the vendor names of the project. Consequently, we needed to investigate the author list of the projects to determine ownership. If the author is affiliated with the library organization, we may consider it a library reuse.

Note that the confirmation of the ground truth is sound but not complete, and not all library reuses may have been identified. More details are specified in \Cref{sec:threat}. However, we assert that all confirmed library reuses are accurate. Out of the 1000 projects, 68 of them have been confirmed to have reused TPLs, resulting in a total identification of 167 TPLs. 

\subsection{RQ1: Effectiveness Evaluation}

\noindent\textbf{Tool Preparation}\label{sec:toolprepare}
CENTRIS~\cite{woo2021centris} is a state-of-the-art clone-based SCA tool. Although the authors primarily focused on C/C++ GitHub projects, they claimed that CENTRIS is not restricted to a particular language. Unlike class-level clones, CENTRIS operates on function-level clones, which makes it a suitable tool for comparison. However, to adapt it to Java, certain modifications are necessary. Originally, CENTRIS processes GitHub projects and queries the tags for versions and release timestamps. It utilizes Universal Ctags~\cite{ctags} to extract function details from C/C++ files. Fortunately, Universal Ctags is able to handle multiple languages including Java. Therefore, we made the following two minor modifications: \ding{172} we modified its input to accept local source code folders with versions instead of GitHub projects. \ding{173} We configured the command to invoke Universal Ctags for analyzing Java source code files. The core algorithms and business logic remained unchanged. To distinguish this modified version from the original, we named it CENTRIS4J.

\noindent\textbf{Experiment Setups}
In this RQ, we first collected the class features from the \textit{Reference Dataset}, following the procedures in \Cref{sec:methodology}. We obtained 607,847 unique features in total. Next, we assess both \toolname and CENTRIS4J on \textit{Test Data with Ground Truth} with precision, recall, and F1-score. Moreover, according to CENTRIS~\cite{woo2021centris}, threshold setting have a great influence on the performance of CENTRIS. Thus, wWe invested considerable effort in adjusting its threshold from 5\% to 25\%, observing a gradual decrease in recall from 0.730 to 0.497, finding that a 15\% threshold offers optimal performance in its F1-score. The following experiment of CENTRIS4J is then conducted with the threshold of 15\%.

\begin{table}
\small
    \caption{Effectiveness Evaluation Results}
    \label{tab:effective}
    \centering
    \begin{threeparttable}
    \scalebox{1}{\begin{tabular}{l|ccc}
    \hline
     \textbf{Name} & \textbf{Precision}     & \textbf{Recall}       & \textbf{F1-score}     \\
    \hline
    \toolname & 0.698 & 0.986 & 0.818 \\
    CENTRIS4J & 0.270 & 0.706 & 0.391 \\
    \hline
    \end{tabular}}
    \end{threeparttable}
\end{table}

\noindent\textbf{Results}
As shown in \Cref{tab:effective}, \toolname outperforms CENTRIS4J. Specifically, \toolname demonstrates superior precision over CENTRIS4J, with scores of 0.698 compared to 0.270, respectively. Additionally, \toolname's recall rate surpasses that of CENTRIS4J, with values of 0.986 compared to 0.706. Consequently, this leads to \toolname achieving a higher F1 score than CENTRIS4J, at 0.818 compared to 0.391, highlighting its enhanced effectiveness.

\noindent \textbf{Why \toolname performs better:} \lida{Compared to \toolname, CENTRIS4J exhibits a lower recall rate. Beyond its inability to maintain inter-function relations, CENTRIS4J's reliance on a fixed threshold further limits its effectiveness. CENTRIS determines TPL reuses only if the matched functions exceed a threshold. However, this threshold is subjective and varies with the range of TPL functionalities and extent of cloning. In Java, TPL reuse often involves more precise code copying, frequently below the threshold. The precision suffers due to this fixed threshold as TPLs are cloned differently depending on their functionalities, making a constant threshold inadequate for reflecting TPL reuse. In our method, we discarded the fixed threshold strategy in TPL recognition, focusing instead on identifying important classes.} 

\begin{figure}
    \centering
    \includegraphics[width=0.7\textwidth]{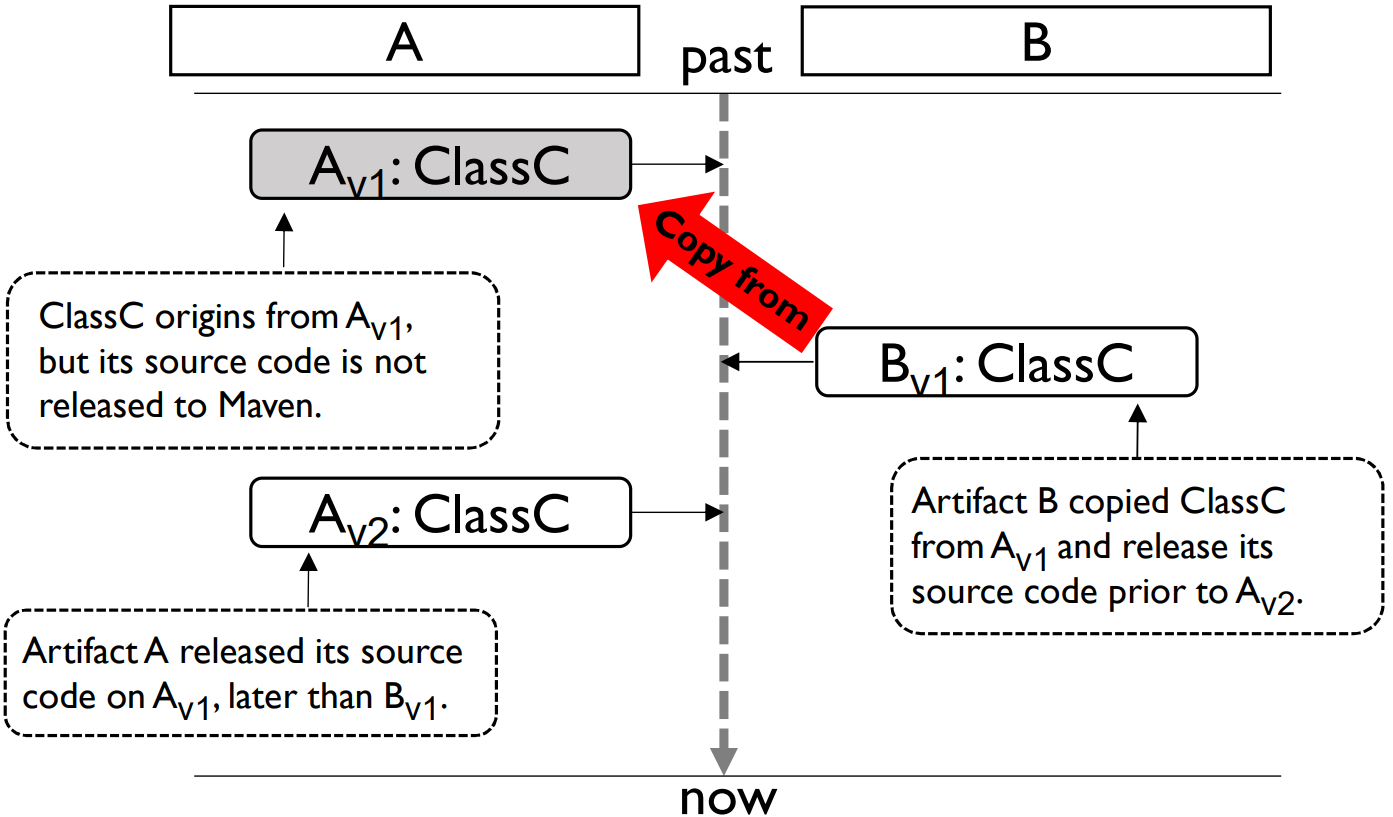}
    \caption{Clone Timeline Example}
    \label{fig:clone_timeline}
\end{figure}

\noindent \textbf{What could impair \toolname:} The factor contributing to the impaired recall of \toolname is the absence of Maven source code data. Not all artifacts have their source code released. Out of the 543,286 versions of the 9,965 libraries in the \textit{Reference Dataset}, 158,316 versions do not have corresponding source code JARs. Some artifacts lack older versions of source code releases, while a few artifacts do not provide any source code JARs on Maven Central at all. \Cref{fig:clone_timeline} illustrates how missing source code data can contribute to the reduced recall of the tool. Considering two artifacts \textit{A} and \textit{B}, both of them have a class namely \textit{ClassC} which is origin from version 1 of artifact \textit{A} ($A_{v1}$). However, $A_{v1}$ is not released with its source code JAR. After that, version 1 of artifact \textit{B} ($B_{v1}$) is released with its source code and it copies \textit{ClassC} from $A_{v1}$. Finally, artifact \textit{A} releases its source code in $A_{v2}$ after $B_{v1}$. As a consequence, during the process of filtering classes by clones in \Cref{sec:filterbyclone}, \textit{ClassC} will be incorrectly regarded as originating from artifact \textit{B}, leading to both false positives and false negatives in the analysis. 
Furthermore, we found some classes do not originate from well-known package managers, like MCR, but are widely accepted by the developers. These classes cannot be imported through standard package manager manifest files. Instead, they are independently released on personal code sites or blogs. Popular examples include Base64 (2011)~\cite{base64web}, HTML Filter (2012)~\cite{htmlfiler}, Base64Coder(2010)~\cite{sourcecoderelease}. Some algorithms from papers are also provided by authors, such as the Porter Stemming Algorithm ~\cite{Porter1980}. The paper was published in 1980 and the author released its code for all languages on its main page~\cite{porterweb} in 2006. These web-source libraries are not included in the ground truth and are considered false positives. 
We also observed numerous classes auto-generated by compilers such as JavaCC~\cite{javacc}, and Thrift Compiler~\cite{thrift}. These generated classes usually contain comments to indicate that they are auto-generated. However, code clones are unable to establish a direct connection between the auto-generated code and the generator libraries because such libraries do not include the code generated by themselves. Consequently, the clones are mistakenly linked to other TPLs that also include the auto-generated code, resulting in false positives.


\begin{tcolorbox}[boxrule=0.5pt,arc=1pt,boxsep=-1mm,breakable]
\textbf{Answering RQ1:} 
\toolname achieved a precision of 0.698, a recall of 0.986, and an F1-score of 0.818, outperforming CENTRIS4J in precision, recall, and F1-score by 0.462, 0.353, and 0.475 respectively.
\end{tcolorbox}

\subsection{RQ2: Efficiency Evaluation}
\noindent\textbf{Experiment Setup.}
In this section, our objective is to compare the efficiency of \toolname and CENTRIS4J in feature generation and TPL recognition. For \toolname, feature generation consists of Feature Extracting and Feature Refining, while for CENTRIS4J, it has component DB construction. To measure the efficiency of TPL recognition, we compared their scan time on 1000 projects from the \textit{Test Data with Ground Truth}. 

\begin{table}[]
\caption{Time Comparison}
    \centering
    \begin{tabular}{lcc}
    \toprule
        & \textbf{\toolname} & \textbf{CENTRIS4J} \\
    \midrule
    Feature Generation & 52 h & 216 h\\
    TPL Recognition (per project) & 14.2 s & 126.6 s \\ 
    \bottomrule
    \end{tabular}
    \label{tab:featuregentime}
\end{table}
\begin{figure}
    \centering
    \includegraphics[width=0.8\textwidth]{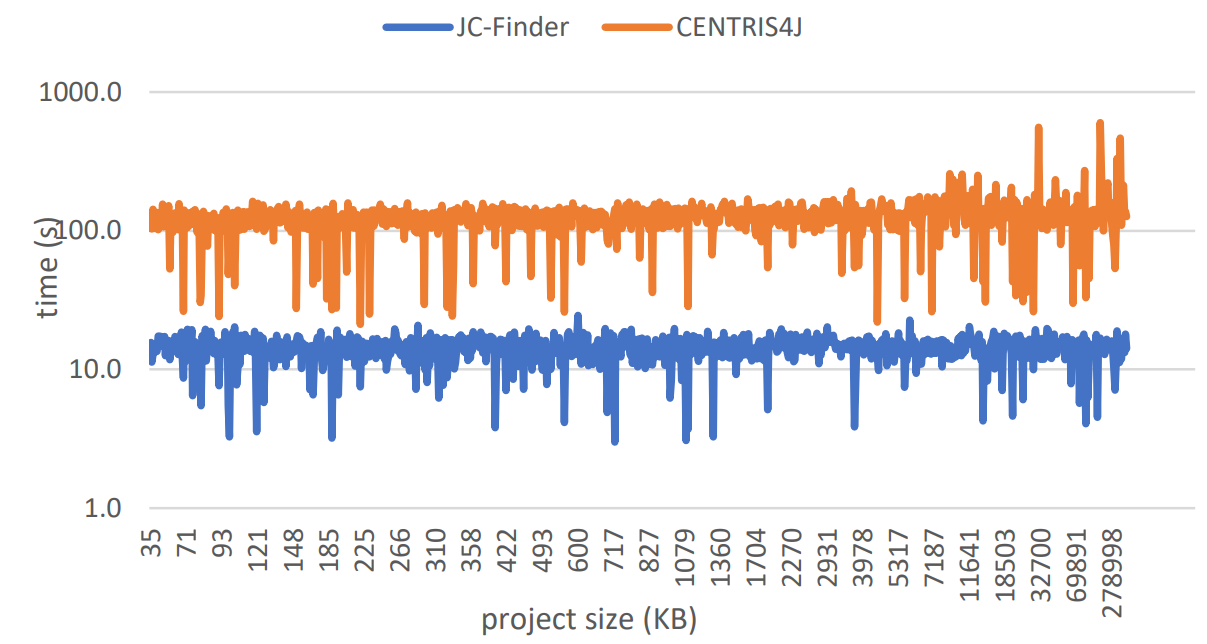}
    \caption{Time Comparison of TPL Recognition}
    \label{fig:tplrecogtime}
\end{figure}

\noindent\textbf{Results.}
\Cref{tab:featuregentime} shows the processing time for feature generation and the TPL recognition time per project. To generate features for 543,286 versions of 9,965 libraries from the \textit{Reference Dataset}, \toolname used 52 hours (around 2.1 days), which is about 4 times faster than CENTRIS4J, which took 220 hours (around 9.2 days). \Cref{fig:tplrecogtime} shows the relation between project size and the scan time for both tools. For both tools, the TPL recognition time does not have a strong positive relation with the size of the project. However, the scan time of \toolname is about nine times faster than that of CENTRIS4J, with an average scan time of 14.2s and 126.6s. 
Also, the scanning time of \toolname is more stable than that of CENTRIS4J with variances of 2.87 and 39.39 respectively. 

\begin{tcolorbox}[boxrule=0.5pt,arc=1pt,boxsep=-1mm,breakable]
\textbf{Answering RQ2:} 
\toolname outperforms CENTRIS4J in efficiency, generating features four times quicker. Furthermore, it is nine times faster and significantly more stable than CENTRIS4J in identifying TPL reuse.
\end{tcolorbox}


\subsection{RQ3: Dependency Detection Enhancement Analysis}
\noindent\textbf{Experiment Setups.} 
This research question aims to compare TPL reuse identified through code cloning with the dependency lists explicitly stated through package managers.

The experiment was conducted on the 7,974 GitHub repositories from the \textit{Test Dataset}. Initially, the \textit{Test Data} was cloned from GitHub, and the user-specified dependencies were parsed from the POM files. These dependencies are referred to as $\mathit{TPL_{PM}}$. Next, the \textit{Test Dataset} were scanned with \toolname to obtain TPLs introduced by code clones, namely $\mathit{TPL_{CC}}$. 
Note that the $\mathit{TPLs_{PM}}$ are imported as Java Archives (JARs), while $\mathit{TPLs_{CC}}$ are imported by copy-and-paste source code. The detection of $\mathit{TPLs_{CC}}$ enhances and supplements the results obtained through $\mathit{TPL_{PM}}$ detection, rather than replacing them. 
In the analysis, we mainly focus on three aspects. \ding{172} How prevalent is $\mathit{TPLs_{CC}}$? \ding{173} How does identifying $\mathit{TPLs_{CC}}$ improve the results compared to only identifying $\mathit{TPLs_{PM}}$? \ding{174} How do $\mathit{TPLs_{CC}}$ and $\mathit{TPLs_{PM}}$ overlap?

\noindent\textbf{Results.}
To answer the first question, among the 7,947 test projects, 789 projects (about 9.89\% of all projects) were found to use $\mathit{TPLs_{CC}}$, and 2,142 $\mathit{TPLs_{CC}}$ were located. Although importing TPLs by package managers is still common practice, the usage of $\mathit{TPLs_{CC}}$ is non-trivial and cannot be ignored. Additionally, we conducted an experiment to explore correlations between the number of $\mathit{TPLs_{CC}}$ and project star numbers, as well as the correlations between the number of $\mathit{TPLs_{CC}}$ and project size, using Pearson correlation~\cite{pearson_wiki} analysis.
The correlation value $r_{xy}$ ranges from -1 to 1, where $0<|r_{xy}|<0.3$ indicates a weak correlation, $0.3<|r_{xy}|<0.6$ indicates a moderate correlation, and $0.8<|r_{xy}|<1$ indicates a strong correlation~\cite{pearson}. The correlation between $\mathit{TPL_{CC}}$ numbers and star numbers is 0.065 and the correlation between $\mathit{TPL_{CC}}$ numbers and project size is 0.232, which indicates that neither project stars nor project size are positively related to the number of $\mathit{TPLs_{CC}}$.

To address the second question, we computed the improvement rate (IR) for each of the 789 projects by determining the percentage of additional $\mathit{TPLs_{CC}}$ discovered with the \toolname in comparison to the initial $\mathit{TPLs_{PM}}$. The IR is calculated by dividing the number of TPLs exclusively identified as $\mathit{TPLs_{CC}}$ by the total number of $\mathit{TPLs_{PM}}$.
$$
IR = \frac{\#(TPL_{CC} - TPL_{PM})}{\#(TPL_{PM})} \times 100\%
$$
Note that for projects that do not contain any $\mathit{TPLs_{PM}}$ but have more than one $\mathit{TPLs_{CC}}$, the IR is set to 1. Subsequently, we computed the average IR across all 789 projects, representing the IR of \toolname, which demonstrates a high IR of 26.20\%. According to our observation, the projects may introduce TPL by code clone for four main reasons. 
Firstly, some target algorithms are initially released on platforms other than MCR. For example, the popular version of the Porter Stemming Algorithm was released on a website in 2006~\cite{porterweb}. Even though an official release on MCR already existed~\cite{opennlp}, the developers may be unaware of it or be reluctant to refactor their code accordingly. 
Secondly, the developers only need part of the code. The developers precisely copy the necessary code segments to facilitate its development. In such cases, developers usually mark the originality of the copied code using phrases such as ``This file is part of...'', ``This package is based on the work done by...'', ``This class is a modified version of the...'', ``copy from...'', ``forked from...'', etc. This approach allows for more flexibility and reduces the size of released distributions. 
Thirdly, some of the projects explicitly place all the code of a library under a specific folder. The folder is usually called ``/ext'', ``/shared'', ``/jre'', or the group and artifact name of the library. Such practices are common in C/C++, where no mature and unified package managers exist. Also, copying the entire project enables  the project to function independently of package managers  and be portable. 
Fourthly, we also found some developers copied the code but only modify the license information, including but not  limited to removing the license, and changing the author name. For example, \textit{simioni87/auth\_analyzer} utilizes a class namely \textit{Diff\_match\_patch} whose origin name is \textit{DiffMatchPatch}, and the authors of \textit{hf200012@oceanus.bi} removes all license information of \textit{Base64} and replaces with his own name. The author conceals the notice of code cloning and such changes violate open-source license agreements~\cite{apachelicense}.

To answer the third question, we calculated the duplication rate (DR) by dividing the common TPLs identified in both $\mathit{TPLs_{CC}}$ and $\mathit{TPLs_{PM}}$ by the number of $\mathit{TPLs_{PM}}$:
$$
DR = \frac{\#(TPL_{CC} \bigcap TPL_{PM})}{\#(TPL_{PM})} \times 100\%
$$
The duplication rate is calculated for each of the 789 projects, and the average duplication rate over all 789 projects is 1.08\%. A lower duplication rate suggests that $\mathit{TPLs_{CC}}$ complement $\mathit{TPLs_{PM}}$ more effectively. 
Most of the duplications occur when the libraries are also released to GitHub. For example, \textit{io.vertx:vertx-rx-java-gen} and \textit{io.vertx:vertx-rx-java2-gen} are released from \textit{vert-x3/vertx-rx}~\cite{vertxrx}. 

\begin{tcolorbox}[boxrule=0.5pt,arc=1pt,boxsep=-1mm,breakable]
\textbf{Answering RQ3:} 
\toolname identified an additional 26\% of TPL reuses beyond what was discovered with PM-based TPL reuse results on average. 
\end{tcolorbox}

\section{Discussion}
In this section, we begin by discussing the threats to validity. We then offer recommendations for developers, SCA users, and open-source library producers. 

\subsection{Threats To Validity}\label{sec:threat}

\noindent \textbf{The collection of ground truth is sound but may not be complete.} 
In cases where there is insufficient detail in paths and names, logs, copyright, license information, or author lists, differentiating between TPL reuse and mere code writing similarity becomes challenging, even for experts. To ensure soundness, we ignore such examples.

\noindent \textbf{The accuracy of the results can be influenced by the size and quality of the \textit{Reference Dataset}.} If a library is not present in the \textit{Reference Dataset}, any reuse of it will not be detected. Also, if some versions of an artifact are missing, the timestamp of related features may be incorrect, potentially leading to inaccurate reports of TPL reuse. Therefore, using a large and complete \textit{Reference Dataset} can improve the accuracy of clone detection. In this work, the \textit{Reference Dataset} is derived from MCR, and more library sources could be included to enhance the performance of \toolname.

\noindent \textbf{The determination of the threshold for trivial functions is influenced by the expertise of experts.} People with varying levels of experience might make different judgments regarding this threshold. However, in this work, all the authors agreed that functions with a complexity of less than 60 are trivial.

\subsection{Suggestions}

\noindent \textbf{Dependencies that cannot be ignored in SCA: $\mathit{TPLs_{CC}}$ should always be considered when conducting SCA.} 
According to RQ3, The findings indicate that \toolname improves PM-based SCA by 26.20\% with only 1.08\% of duplications. Therefore, when scanning a project, the users should always pay attention to the source of the TPL and ensure the TPLs imported by code clones are included. Noting that the purpose of \toolname is not to replace the detection of $\mathit{TPL_{PM}}$, instead, it complements the results of PM-based Java SCA and achieves more comprehensive and accurate results.

\noindent \textbf{The reasons of using $\mathit{TPL_{CC}}$ and possible solutions.} As we have shown in RQ3, the number of $\mathit{TPL_{CC}}$ has poor correlation with project stars and size. There are two main reasons why users prefer importing TPLs through code cloning rather than relying on $\mathit{TPL_{PM}}$. Firstly, this approach is adopted when no suitable libraries are available in MCR, making them difficult to locate or utilize effectively. For instance, libraries like Porter Stemmer~\cite{porterweb} and Base64~\cite{base64web} were initially released on websites, and many projects incorporated them. Even though they have been integrated into other MCR libraries, such as opennlp-tools~\cite{opennlp}, users may remain unaware of this fact and continue to use the original versions.
Secondly, code clones in Java are highly precise, often requiring only a portion of the code, especially when dealing with excessively large libraries that encompass numerous independent functionalities. In such cases, users may opt for a partial copy of the library rather than importing the entire package to manage the overall size. One potential solution is to divide certain libraries into smaller, standalone, and more specialized components based on their functionalities. This approach not only makes tracking easier but also reduces the load carried by developers.

\noindent \textbf{Suggestions to developers: it is advised to prefer using $\mathit{TPLs_{PM}}$ over $\mathit{TPLs_{CC}}$.} 
Although importing TPLs by code clone grants greater control over project size and development freedom, it suffers from poor version control, whereas package managers provide better version management based on version numbers. Such weakness could pose a risk of not being updated and keep potential vulnerabilities persist for a long time. In contrast, package managers promptly inform users about new releases, ensuring better security.
Although, in earlier years, some libraries were only available on websites with source code, the trend has shifted towards most libraries being packaged as released libraries with improved interfaces and efficiency. Consequently, developers are encouraged to explore existing libraries before resorting to code cloning.

\section{Conclusion}
We proposed \toolname which is a class-level clone-based SCA tool for Java projects. We then evaluate the performance and efficiency of \toolname and compared it with CENTRIS in the context of Java source projects. \toolname outperforms the other tool in both performance and efficiency. Next, we scan the TPL reuse introduced by code clones on a large group of GitHub projects with \toolname. \toolname is proven to be effective in complementing the SCA that only detects TPLs from package managers by finding 26.20\% more TPLs. 
\section{Data Availability}
The data can be publicly accessed at \url{https://anonymous.4open.science/r/JC-Finder/}.

\balance
\bibliographystyle{ACM-Reference-Format}
\bibliography{ref}

\end{document}